# A Simultaneous Model to measure Academic and Financial Performances of Scientific Activities

L.T. Handoko
Group for Theoretical and Computational Physics, Research Center for Physics,
Indonesian Institute of Sciences, Kompleks Puspiptek Serpong, Tangerang 15314,
Indonesia
handoko@lipi.fisika.net

ABSTRACT :
I propose a new model to measure simultaneously academic and financial performances of scientific activities quantitatively. The tool is very simple and can be applied to any branches of science, while it is also adjustable to varying macro-economic indicators. I argue that implementing the model could realize a fair and objective decision-making and also reward and punishment system in order to improve the individual and institutional performances in scientific activities.

Keywords : economic model, science and technology indicator

## I. INTRODUCTION

Management is an important aspect of human-being and its activities, including any kind of scientific activities. Management as the regulator and also the executor of regulation needs appropriate tools to implement the existing regulation. Regulation has been created as a common norm and commitment which should be followed by all (individual and institutional) fellows in order to establish a mutual relationship. Mutual relationship is a crucial point to realize every fellow's purposes and motivate all of them to work together.

In the context of scientific activity, scientific management plays an important role, much more than the other non-scientific (bureaucracy, business, etc) activities. This reflects the nature of scientific activity which is strongly based on the individual with unlimited degree of freedom. Because independency and freedom are necessary conditions for any scientific activities. On the other hand since a scientific activity is mostly supported by public fund, all scientific fellows and its activities should be accountable for public. In a real life, however fully transparent, easily-understandable and accountable scientific activities are often difficult to be realized. This is moreless due to the nature of scientific activity which is in most cases invisible and unpredictable. Fortunately, in contrast with another non-scientific activities, scientific activity is always supposed to generate an objective and calculable outcomes in a period of duration.

This means there is an urgent need for a specific tool to measure scientific performance based on the scientific outcomes. Because scientific outcome is the only part which is measureable. Moreover, performance measurement is neither limited to scientific performance nor long term economic potential. So far, scientific activities are almost justified by the aspect of economic potential using the method of cost-benefit analysis [1]. While the scientific performance is measured naively based on the scientific outcomes. This method clearly separates scientific and financial aspects, although both aspects are closely related each other. More than that, in daily practice it is hard to measure the objectivity on cost-benefit method since it uses absurd references as future potential which is yet unpredictable.

It is also known more complete method as scientific and technical human capital (STHC) [2] which observes a lot of aspects. Though the method looks ideal, this model is very complicated and involves a lot of subjective parameters. This reduces the accuracy and validity of the method as an easy, transparent and obejective measurement tool.

## II. THEORETICAL FRAMEWORK

According the introduction in the preceeding part, it is clear that as long as concerning scientific activity, performance should be measured based on the achievement of scientific outcome without considering its process. Inversely, this point makes scientific activity is easy to measured and quantized. Motivated by this fact, we propose an alternative tool called as OCSP Model [3].

As mentioned above, the first assumption in the model is all measurements are based only on scientific outcomes regardless its process. The scientific outcome itself is defined as : all outcomes generated in a scientific activity which have been approved by independent third party(ies) in a form of either scientific documents or other real activities.

The second assumption, the measurement is done in a year basis and takes into account all outcomes in the last fiscal year.

Next assumption is, each outcome is ranked based on its difficulty to achieve, and is provided by an appropriate point based on its scientific content. The order number ($N_O$) of all relevant scientific outcomes must be in order and not doubled. On the other hand, its scientific point ($S_P$) may be the same with another neighboring scientific outcomes, but it must be smaller (greater) than another scientific outcomes with different points above (below). Determination of the order and the point of scientific outcomes may differ depend on the nature of each field of science.

Further, it is also assumed some parameters of maximum scientific point ($P_M$), descending rate of scientific point ($P_D$, in percent) and total scientific point treshold per-scientist ($P_T$). These parameters are the same of all fields of science. This method enables universal evaluation and comparison among different branches of science. It should be remarked that the absolute value of scientific point itself is less important, because it only represents the scale of differences among scientific outcomes. Once the maximum scientific point and its descending rate have been determined, a group of available scientific points for each scientific outcomes can be obtained as :

$$S_P = \{P_M, P_D \times P_M, (P_D)^2 \times P_M, ..., (P_D)^{no-1} \times P_M, (P_D)^{no} \times P_M\} \quad (1)$$

where $n_O$ denotes the number of relevant scientific outcomes in a field. Concerning the appropriateness, it is natural to restrict the values for following parameters :

$$\text{minimum } S_P > 1, \quad (2)$$
$$P_M, P_T > 100, \quad (3)$$
$$90\% < P_D < 50\% \quad (4)$$

Economic aspect related to financial performance is represented by the economic coefficient ($C_E$) which holds for all branches of science. This parameter should be determined initially, and thereafter can be made varying automatically depending on the macro economic indicators (inflation, economic growth, currency rate, etc).

Finally, using all assumptions above the reatio of Scientific Performance (*SP*) and Financial Performance (*FP*) are formulated as the following equations. From these equations, it is clear that *FP* is related directly to the scientific point and order number of each scientific outcome.

$$SP = \frac{1}{n_P \times P_T} \left[ \sum_{i=1}^{n_o} (S_P)_i \times (Q_o)_i \right] \qquad FP = \frac{C_E}{B_T} \left[ \sum_{i=1}^{n_o} \frac{(S_P)_i \times (Q_o)_i}{(N_O)_i} \right] \quad (5)$$

where $n_P$ : number of scientists involved in an activity, $Q_O$ : quantity of a scientific outcome and $B_T$ denotes total budget in a fiscal year.

### III. APPLICATION

Regarding the research management, in the evaluation based on OCSP Model it is required a regulation in two levels of management to determine all parameters introduced above. That is :

1. <u>Official bureaucrat with authority on scientific activity</u>.

    The management in this level is responsible for determining the values of main parameters which hold for all branches of scientific activities, i.e. $P_M$, $P_D$, $P_T$ and $C_E$.

2. <u>Related scientific community</u>.

    The scientific community in a specific field is responsible for determining the order

and the value of scientific point of all relevant scientific outcomes. Because these parameters are unique for a specific field.

Soon after determining all parameters. the system is ready for evaluating previous scientific activities and also decision-making of future scientific activities as well. As an example one can use an online facility available on the net [4].

## IV.CONCLUSION AND DISCUSSION

Using the OCSP Model, a simultaneous relation between scientific and financial performances has been formulated based only on the scientific outcomes. Several advantages of measurement tool based on the OCSP Model :

- Since it is based only on the scientific outcomes, the obejectivity and transparency of measurement can be guaranteed.
- A single year base method makes the evaluation process and decision-making for next fiscal year easier, since the result reflects an up-to-date real condition.
- In long term, all annual evaluations of eaach fellow can be compiled to implement a compensation system. Example : distribution and allocation of human resources and funding to each individual or institution.
- The same tool can be used to evaluate incoming proposals regarding the promising targets claimed in.
- The result of financial convertion can be used as a standard to decide the budget approved for each individual or scientific project.

Implementing the OCSP measurement tool consistently and continously in long term for each individual, scientific activity and institution could provide performance indicators in each level. This has a great potential to help identifying the fundamental problems in regional or national scale. Combining the result with a strict compensation system could also improve the whole scientific performance.

The parameters introduced above are not absolute and only indicate the scale, which make them to be determined easily without requiring deep consideration. However, more serious consideration is still required for few of them, which especially constitute all fields absolutely as descending rate of scientific point ($P_D$) and total scientific point treshold per-scientist ($P_T$). The parameter PT can be determined in a comprehensive way by for instance taking the mean of all scientific outcomes generated by scientists around the world. The data can be retrieved easily through the existing databases available on the net. This result further can be used to determine the appropriate mean of parameter $P_D$. Therefore, further research utilizing available databases providing global and local scientific outcomes on the net is very recommended.